\documentclass[a4paper,twocolumn,superscriptaddress,11pt,accepted=2017-05-09]{quantumarticle}
\pdfoutput=1
\usepackage[utf8]{inputenc}
\usepackage[english]{babel}
\usepackage[T1]{fontenc}
\usepackage{amsmath}
\usepackage{hyperref}
\usepackage{amsthm}
\usepackage{bbm}
\usepackage{graphicx}
\usepackage{tikz}
\usepackage{lipsum}

\usepackage[numbers,sort&compress]{natbib}

\def\C{{\mathbbm C}}

\def\id{{\mathbbm I}}
\def\H{{\cal H}}
\def\tr{\mbox{tr}}
\def\sp{\mbox{span}}
\def\Q{{\cal Q}}
\def\R{{\cal R}}
\def\M{{\cal M}}

\def\S{{\cal S}}

\def\be{\begin{equation}}
\def\ee{\end{equation}}
\def\tr{\mbox{tr}}
\def\bra#1{\langle#1|}
\def\ket#1{|#1\rangle}
\def\braket#1#2{\langle#1|#2\rangle}

\def\proj#1{\ket{#1}\!\bra{#1}}

\newtheorem{theo}{Theorem}
\newtheorem{prop}[theo]{Proposition}

\begin{document}

\title{Bond dimension witnesses and the structure of homogeneous matrix product states}
\date{\today}
\author{Miguel Navascu\'es}
\affiliation{Department of Physics, Bilkent University, Ankara 06800, Turkey}
\affiliation{Institute for Quantum Optics and Quantum Information (IQOQI), Boltzmangasse 3, 1090 Vienna, Austria}
\author{Tam\'as V\'ertesi}

\affiliation{Institute for Nuclear Research, Hungarian Academy of Sciences, H-4001 Debrecen, P.O. Box 51, Hungary}
\maketitle

\begin{abstract}
For the past twenty years, Matrix Product States (MPS) have been widely used in solid state physics to approximate the ground state of one-dimensional spin chains. In this paper, we study homogeneous MPS (hMPS), or MPS constructed via site-independent tensors and a boundary condition. Exploiting a connection with the theory of matrix algebras, we derive two structural properties shared by all hMPS, namely: a) there exist local operators which annihilate all hMPS of a given bond dimension; and b) there exist local operators which, when applied over any hMPS of a given bond dimension, decouple (cut) the particles where they act from the spin chain while at the same time join (glue) the two loose ends back again into a hMPS. Armed with these tools, we show how to systematically derive `bond dimension witnesses', or 2-local operators whose expectation value allows us to lower bound the bond dimension of the underlying hMPS. We extend some of these results to the ansatz of Projected Entangled Pairs States (PEPS). As a bonus, we use our insight on the structure of hMPS to: a) derive some theoretical limitations on the use of hMPS and hPEPS for ground state energy computations; b) show how to decrease the complexity and boost the speed of convergence of the semidefinite programming hierarchies described in [Phys. Rev. Lett. 115, 020501 (2015)] for the characterization of finite-dimensional quantum correlations.
\end{abstract}

\section{Introduction}

The study of condensed matter phases depends crucially on our ability to determine the properties of the ground state of local Hamiltonians defined over a lattice. Not only this problem has been shown to be QMA-hard for general Hamiltonians \cite{complex1,complex2}, but already for mesoscopic systems (of $n\sim 40$ particles), even \emph{storing} the description of a general quantum state in a normal computer becomes an impossible task. This forces condensed matter physicists to resort to quantum state ansatzs in order to understand and study the properties of matter in the low temperature regime. 

One ansatz that has proven very useful in this respect is the family of Tensor Network States (TNS) \cite{tensor_net}, a class of many-body wavefunctions of complexity fixed by a parameter known as \emph{bond dimension}. In the last few years, TNS have been successfully used to approximate the low energy sector of local Hamiltonians of spin lattices of different dimensions \cite{app_MPS,peps,MERAS}. The ability to approximately compute expectation values in an efficient manner, together with the possibility to conduct optimizations in the thermodynamical limit \cite{iMPS1}, \cite{iMPS2}, \cite{iPEPS}, \cite{iPEPS2} makes TNS one of the very few avenues to understand the physics of strongly correlated systems. 

Because of all the above and further theoretical considerations, in the last years, almost every talk about TNS starts with the speaker reminding the audience that `TNS of low bond dimension are the \emph{only} physical states of condensed matter systems', or, equivalently, that `all other rays of the Hilbert space of a many body system are not physically realizable'. 

Suppose that we take this last claim at face value. That is, we postulate that the laws of Nature are such that the states of condensed matter systems at low temperature are very close to being representable by convex combinations TNS of low bond dimension. This is in effect a physical theory; as such, its limits must be explored to determine to which degree the theory is falsifiable, and thus scientific. Given that all one can hope to estimate in the lab are the expectation values of certain $k$-local observables, what makes TNS special, when compared to any other quantum state? Can we prove that the underlying quantum state cannot possibly be approximated by convex combinations of TNS with low bond dimension, i.e., can we falsify a \emph{TNS model}? Note that, even if we do not adhere to the ``Church of the Low Bond Dimension", we can use the bond dimension of a TNS model as a measure of the complexity of the underlying quantum state. In this sense, an experimental refutation of a TNS model with high bond dimension could be regarded as a benchmark for the quantum control of condensed matter systems.

We lack tools to answer these questions. Note that the naive scheme of lower bounding the bond dimension by estimating the rank of reduced density matrices only works with pure TNS and not convex combinations thereof. Moreover, the physical scenarios which we consider here just allow the experimentalist to estimate averages of two-body reduced density matrices. In addition, the variational methods so commonly used in condensed matter physics to optimize over TNS of fixed bond dimension are useless to refute a TNS model: such a task would require \emph{relaxation}, rather than variational techniques.

In this paper, we will address these problems for homogeneous Matrix Product States (hMPS) \cite{mps}, a class of TNS used to model non-critical one-dimensional spin chains with translation invariance. 

We start by deriving two surprising features of hMPS: first, for any $D$ we identify local operators which annihilate all hMPS of bond dimension smaller than or equal to $D$. Second, for any $D$ we prove the existence of local operators which, when applied over any hMPS of bond dimension $D$, decouple (cut) the particles where they act from the spin chain while at the same time join (glue) the two loose ends back again into a hMPS.

Armed with these notions, we will define a family of $k$-local operators with negative eigenvalues whose expectation values are nonetheless positive for all hMPS of a given bond dimension $D$. Each such operator can be used to certify that the quantum state of a non-critical spin chain does not admit an hMPS representation of bond dimension $D$. Moreover, this partial characterization of the dual cone of hMPS allows us to devise general \emph{feasibility tests}, or automated criteria to falsify hMPS models given limited data about the underlying quantum state, such as a number of experimentally available expectation values.

In addition, we will construct instances of local Hamiltonians of arbitrarily many qubits for which a blind application of hMPS-based optimization methods would fail to estimate the ground state energy. This construction can be generalized to other TNS for optimizations over spin lattices of higher spatial dimensions. We will also exploit the low dimensionality of the space spanned by hMPS and the notion of cut-and-glue operators to decrease the complexity and boost the speed of convergence of the semidefinite programming (SDP) \cite{sdp} hierarchies described in \cite{fin_dim,fin_dim_long} for the characterization of finite-dimensional quantum correlations.

The structure of this paper is as follows: first, we will introduce local Hamiltonians and MPS, and also a couple of notions from the theory of Matrix Algebras. Then we will reveal a connection between hMPS and polynomials of non-commuting variables, which will allow us to derive non-trivial structural properties of hMPS. Next, we will use these properties to explore the limits of hMPS models and improve the SDP relaxations proposed in \cite{fin_dim,fin_dim_long}. Finally, we will discuss how some of our results generalize to Projected Entangled Pairs States (PEPS) \cite{peps}.

Before we proceed, though, a disclaimer is in order: long after the completion of this work, we were made aware that the connection between matrix algebras and MPS had already been pointed out by R. Werner in 2006 \cite{Werner_en}. In this encyclopaedia article, Werner also observes that the dimensionality of the space spanned by MPS of a fixed bond dimension is polynomial on the system size.

\section{Matrix product states and the theory of matrix algebras}

Consider a quantum system composed of $n$ distinguishable particles, each of which has local dimension $d$. A general pure state $\ket{\psi}$ of this ensemble hence lives in $(\C^d)^{\otimes n}$, and so we require $d^n$ complex parameters to describe it.

An $n$-site \emph{homogeneous Matrix Product State} (hMPS) is a state of the form

\be
\ket{\psi(\omega,A,n)}=\sum_{i_1,...,i_n=1}^{d}\tr(\omega A_{i_1}A_{i_2}...A_{i_n})\ket{i_1,...,i_n},
\label{MPS}
\ee

\noindent where $\omega,A_1,...,A_{d}$ are $D\times D$ matrices. Note that in general Matrix Product States (MPS) the matrices $\{A_i\}_{i=1}^{d}$ are taken to be site-dependent \cite{mps}. The matrix $\omega$ is a boundary condition, while the parameter $D$ is known as the \emph{bond dimension} of the state. In order to distinguish it from $d$, the latter is also called the \emph{physical dimension}.

It can be proven that, for $D$ high enough, all states can be expressed as in eq. (\ref{MPS}). However, we will be interested in systems where the value of $D$ does not grow much with the system size $n$. Note that a hMPS of whatever size can be described with $O(dD^2)$ complex parameters. Hence, as long as $D$ is not very big, it pays to use this approximation. Finally, notice that, taking $\omega=\id_D$, the state becomes invariant with respect to the permutation $1\to 2\to...\to n\to 1$. Such states are called \emph{uniform translation invariant (TI) MPS}.

For low values of $D$, computing expectation values of product operators in a hMPS can be carried out in an efficient way. Thus hMPS (in particular, uniform TI MPS) are regularly used to approximate the ground state of $k$-local Hamiltonians of one-dimensional systems, i.e., Hamiltonians of the form:

\be
H=\sum_{j=1}^n h_{j},
\label{mark}
\ee

\noindent where $h_{j}$ acts non-trivially on the space of the particles $j,j+1,...,j+k-1$. Given $H$, we will denote by $\langle H\rangle_D$ the minimum average value of $H$ achievable with hMPS of bond dimension $D$.

Note that we can always choose $\{A_i\}$ satisfying $\sum_{i=1}^{d}A_iA_i^\dagger =\id_D$ \cite{mps}. This allows us to perform calculations in the thermodynamic limit, i.e., $n\to\infty$. Indeed, in such a case, the $m$-site reduced density matrix $\rho_{m}$ of the state under consideration is equal to

\be
\rho_m=\sum_{\vec{i},\vec{j}}\tr(A_{j_m}^\dagger ...A^\dagger_{j_1}\sigma A_{i_1}...A_{i_m})\ket{i_1,...,i_m}\bra{j_1,...j_m},
\label{TI_MPS}
\ee

\noindent where $\sigma\geq 0$ satisfies $\tr(\sigma)=1$ and $\sum_{i} A_i^\dagger \sigma A_i=\sigma$, and the sum runs over all vectors $\vec{i},\vec{j}\in\{1,...,d\}^m$. With the latter conditions, the above is an \emph{infinite MPS} (iMPS).

If the state of a finite spin chain can be expressed as a convex combination of hMPS of bond dimension $D$, we will say that it admits a \emph{hMPS model} of bond dimension $D$. Furthermore, if the TI state of an infinite spin chain can be expressed as a convex combination of iMPS of bond dimension $D$, we will say that it admits an \emph{iMPS model}. 

An important notion in uniform MPS is the concept of \emph{injectivity}. A uniform $n$-site TI MPS $\ket{\psi}$ with bond dimension $D$ is said to be \emph{injective} if, for $m\leq n$, the map $\Gamma: B(\C^D)\to\C^{d^m}$ given by $\Gamma(X)=\sum_{i_1,...,i_m}\tr(XA_{i_1}...A_{i_m})\ket{i_1,...,i_n}$ is injective. Equivalently, a uniform TI MPS $\ket{\psi}$ with bond dimension $D$ is \emph{not} injective if it admits a representation in terms of $D\times D$ block-diagonal matrices.

We now digress momentarily from the topic of MPS to the theory of matrix algebras. A matrix polynomial identity (MPI) $F(X)$ for dimension $D$ is a polynomial of noncommuting variables $X_1,...,X_d$ that vanishes when evaluated with matrices of dimension $D$ or lower. For example, any commutator $[X_i,X_j]$ is a MPI for $D=1$. A more elaborate example is $[[X_1,X_2]^2,X_3]$; this polynomial vanishes when evaluated with $2\times 2$ matrices.

Let us see why: being a commutator, the trace of $Z\equiv [X_1,X_2]$ must be zero, and so $Z=\sum_{i=1}^3c_i\sigma_i$, for some complex numbers $c_1,c_2,c_3$ (here $\sigma_1,\sigma_2,\sigma_3$ denote the Pauli matrices). Squaring $Z$ we get $Z^2=(\sum_{i=1}^3c_i^2)\cdot\id$, and thus $[Z^2,X_3]=0$ for all $X_3\in B(\C^2)$.

MPIs do not only exist for dimensions $1$ and $2$. In general, it can be proven that any $2D$-tuple of $D\times D$ matrices $X_1,...,X_{2D}$ must satisfy the \emph{standard identity} $F_{2D}(X)=0$ \cite{MPI}, where

\be
F_N(X)\equiv\sum_{\pi\in S_{N}}\mbox{sgn}(\pi)X_{\pi(1)}...X_{\pi(N)}.
\label{funda}
\ee

\noindent Here $S_{N}$ denotes the set of all permutations $\pi$ of $N$ elements. It can also be shown that MPIs for matrices of dimension $D$ must necessarily have degree $2D$ or higher \cite{MPI}.

A concept related to MPIs is that of \emph{central matrix polynomials}, or polynomials $P(X)$ of noncommuting variables which are proportional to the identity when evaluated with $D\times D$ matrices. E.g.: in $D=1$, any polynomial can be interpreted as a central polynomial. In $D=2$ we already saw an example, namely the polynomial $[X_1,X_2]^2$. As with MPIs, it can be proven that non-trivial central polynomials (i.e., central polynomials which are not MPIs) exist for all dimensions $D$ \cite{MPI}.

\section{The Physics of MPS}

We will next establish a relation between matrix polynomials and many-body quantum states. From this link, non-trivial structural properties of hMPS will follow almost straightforwardly. Let $X\equiv (X_1,...,X_d)$ be any tuple of $d$ noncommuting variables, and let $P(X)$ be a homogeneous polynomial $P(X_1,...,X_d)=\sum_{i_1,...i_m}p_{i_1,...,i_m}X_{i_1}...X_{i_m}$ of degree $m$. By $\ket{P(X)}$ we will denote the $m$-particle vector $\ket{P(X)}=\sum_{i_1,...i_m}p^*_{i_1,...,i_m}\ket{i_1,...,i_m}$. 

It is immediate that, for any $n$-site hMPS $\ket{\psi}$ of the form (\ref{MPS}), applying $\bra{P(X)}$ over particles $s+1,...,s+m$ leads to
\small

\begin{align}
&\braket{P(X)}{\psi}=\nonumber\\
&=\sum_{i_1,...,i_s,i_{s+m+1},...,i_n}\tr(\omega A_{i_1}...A_{i_s}P(A)A_{i_{s+m+1}}...A_{i_n})\times\nonumber\\
&\times\ket{i_1,...,i_s,i_{s+m+1},...,i_n}.
\label{project}
\end{align}

\normalsize

\noindent That is, we obtain a state similar to an $(n-m)$-site hMPS, but with an `impurity' in the middle, namely the matrix polynomial $P(A)$. This notion of interacting with physical sites in order to engineer operators at the virtual level is actually the main idea behind measurement-based quantum computing on MPS \cite{gross}.

Now, let $P(X)$ be a homogeneous MPI for dimension $D$ of degree $m$. Then, according to (\ref{project}), $\braket{P(X)}{\psi}=0$, i.e., the local operator $\bra{P(X)}$ will have the property of annihilating \emph{any} hMPS with bond dimension $D$ or smaller. Conversely, let $\ket{P(X)}$ be an $m$-particle vector with the property of annihilating all hMPS of bond dimension $D$ or smaller. Due to our freedom in choosing the boundary condition $\omega$, it is easy to see that $P(X)$ must necessarily be a MPI. We have just established that the local space spanned by hMPS is the orthogonal complement of the space of homogeneous MPIs for dimension $D$.

Note that, if $\ket{\psi}$ is a non-injective uniform TI MPS, then $\braket{P(X)}{\psi}=0$ for all MPIs $P(X)$ of dimension $D-1$. This follows from the fact that the diagonal blocks of one of its matrix representations must have size $D-1$ or smaller.

The next question to answer is how big these two spaces are. Denote by $\H^{MPS}_{D,m}$ the $m$-local space spanned by hMPS of bond dimension $D$ of whatever size $n\geq m$, and call $\H^{MPI}_{D,m}$ the orthogonal complement of $\H^{MPS}_{D,m}$. Since MPIs of degree smaller than $2D$ do not exist, we have that, for $m<2D$, $\H^{MPI}_{D,m}=\{0\}$. Now, it is easy to see that $\H^{MPS}_{D,m}$ corresponds to $\sp \{\ket{\psi(\omega,A,m)}:\omega,A_1,...,A_d\in B(\C^D)\}$. Calling $\vec{a}$ the entries of the matrices $A$, we thus have that

\be
\ket{\psi(\omega,A,m)}=\sum_{w}w(\vec{a},\omega_{ij},m)\ket{\phi_w},
\label{mon_dec}
\ee

\noindent where the sum runs over all monomials $w$ of degree $m$ in $\vec{a}$ and degree $1$ on the entries of $\omega$, and the vectors $\{\ket{\phi_w}\}_w$ do not depend on the particular values of $A,\omega$ (e.g.: the vector corresponding to the monomial $\omega_{11}(A_1)_{11}^m$ is $\ket{1}^{\otimes m}$). The above decomposition allows us to bound the dimensionality of $\H^{MPS}_{D,m}$ simply by counting the number of such monomials. The result is

\be
\mbox{dim}(\H^{MPS}_{D,m})\leq D^2\left(\begin{array}{c}m+dD^2-1\\dD^2-1\end{array}\right),
\label{UB}
\ee

\noindent where the $D^2$ factor stems from the number of entries of the boundary condition $\omega$. Relation (\ref{UB}) implies that the dimension of $\H^{MPS}_{D,m}$ increases polynomially with the system size $m$, contrarily to the total local space dimension, which increases as $d^m$. In the limit of high $m$, the space of MPIs is therefore exponentially bigger than $\H^{MPS}_{D,m}$. 

Both the identification of $\H^{MPS}_{D,m}$ with the orthogonal complement of $\H^{MPI}_{D,m}$ and the polynomial bound on $\mbox{dim}(\H^{MPS}_{D,m})$ appear in Werner's encyclopaedic article \cite{Werner_en}.

In Appendix \ref{MPS_basis} we describe two efficient (i.e., with time complexity polynomial on $m$) algorithmic procedures to generate an orthonormal basis for $\H^{MPS}_{D,m}$. This allows us to ascertain the exact dimensionality of the spaces $\H^{MPS}_{D,m}$, for whatever values of $D,d,m$. Some results are presented in Table~\ref{MPS_dim}.

\begin{table}
\resizebox{\columnwidth}{!}{\begin{tabular}{|c|c|c|c|c|c|c|c|c|c|c|c|}
\hline
$m$& 5& 6& 7& 8&9&10&11&12&13&14&15\\
\hline
$D=2$& 30& 53& 88& 139&210&306&432&594&798&1051&1360\\
$D=3$& 32& 64& 128& 256&506&976&1820&3278&5700&9597&$\times$\\
$D=4$& 32& 64& 128& 256&512&1024&2048&3278&8192& $\times$&$\times$\\
$D=\infty$& 32& 64& 128& 256&512&1024&2048&4096&8192&16384&32768\\
  \hline
\end{tabular}}
\caption{Dimension of the $m$-qubit subspace $\H^{MPS}_{D,m}$ for different values of the bond dimension $D$.}
\label{MPS_dim}
\end{table}

Let $h\in B(\H^{MPI}_{D,m})$ be a self-adjoint operator acting on the space of MPIs of dimension $D$, and suppose that we integrate it in a $k$-local Hamiltonian $H$, with $k\geq m$. That is, suppose that the Hamiltonian of the system is $H'=H+h$, with $H$ given by eq. (\ref{mark}).

The above discussion implies that hMPS of bond dimension $D$ or lower `will not see' such a term, i.e., $H'\ket{\psi}=H\ket{\psi}$ for all hMPS $\ket{\psi}=\ket{\psi(\omega,A,n)}$, with $\omega,A\subset B(\C^D)$, $n\geq m$. Elaborating on this, we find a limitation common to all hMPS-based variational methods for Hamiltonian minimization:

\begin{prop}
\label{method_fails}
Let $D>D'>1$ be natural numbers. Then, for sufficiently large $n$, there exists a $O(D^2)$-local TI $n$-qubit Hamiltonian $H$ satisfying

\begin{align}
&\langle H\rangle_{D'-1}>\langle H\rangle_{D'}=\langle H\rangle_{D'+1}=...\nonumber\\
&...=\langle H\rangle_{D-1}=\langle H\rangle_{D}> \langle H\rangle_{D+1}.
\label{curioso}
\end{align}

\noindent The result also holds when we restrict the Hamiltonian minimization to uniform TI MPS (with $\omega=\id_D$). Moreover, it can be extended to iMPS and TI Hamiltonians in the infinite spin chain. The Hamiltonian can also be taken $O(D)$-local at the cost of increasing the physical dimension of the particles. 

\end{prop}
\noindent See Appendix \ref{proof_prop1} for a proof.

A blind application of the proverbial method of minimizing a Hamiltonian via hMPS of increasing bond dimension until the sequence of energy values \emph{appears} to converge hence risks getting stuck at a suboptimal point. Admittedly, the limitations implied by the Proposition do not pose a practical threat for usual studies of one-dimensional non-critical chains, since current hMPS-based algorithms allow reaching bond dimensions of order 100 in a normal computer (way beyond the locality of Hamiltonians of physical interest). This is no longer the case, though, for some other classes of TNS, for which we lack good optimization schemes and Proposition \ref{method_fails} also extends, see below. Note though that none of the arguments above apply to MPS or TNS with site-dependent tensors.

Let us conclude with a speculative thought: consider the evolution of hMPS under homogeneous Matrix Product Operators (MPO) rather than under local Hamiltonians. Since MPOs represent polynomials of arbitrarily high degree, in principle there could exist homogeneous MPOs with low bond dimension which nonetheless annihilate all hMPS of a moderately high bond dimension. That could jeopardize the performance of current algorithms to simulate time evolution in condensed matter systems, see \cite{MPO}.

Suppose now that we choose $P(X)$ in eq. (\ref{project}) to be a central polynomial for dimension $D$. Then $P(A)=p(A)\id$, where $p(A)$ is a scalar. The state $\ket{\psi}$ will hence get projected into the state $p(A)\ket{\psi'}$, with

\begin{align}
\ket{\psi'}=\sum_{\begin{array}{c}i_1,...,i_s,\\i_{s+m+1},...,i_n\end{array}}&\tr(\omega A_{i_1}...A_{i_s}A_{i_{s+m+1}}...A_{i_n})\times\nonumber\\
&\times\ket{i_1,...,i_s,i_{s+m+1},...,i_n}.
\label{glued}
\end{align}

\noindent This is again a hMPS with the same boundary condition $\omega$ and matrices $A_1,...,A_d$, but where $m$ particles are just missing. 

\begin{figure}
  \centering
  \includegraphics[width=8 cm]{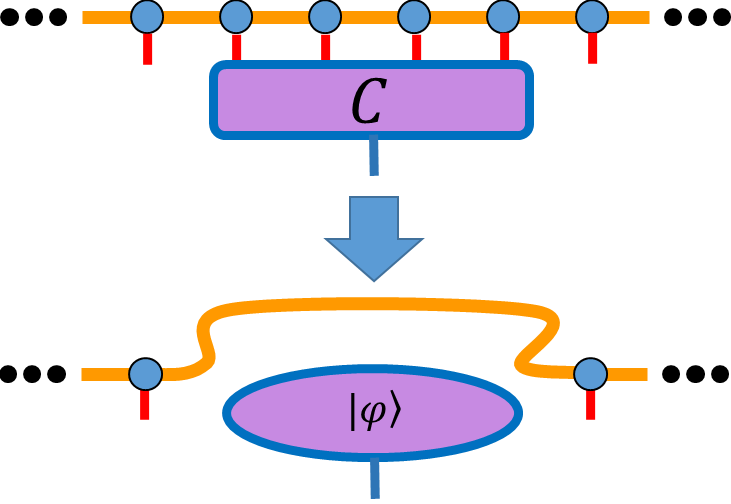}
  \caption{Action of a cut-and-glue operator $C$ over a MPS.}
  \label{cut_and_glue}
\end{figure}

Now, divide the space of $m$-degree homogeneous central polynomials into classes $[P]=P+MPI$, i.e., two central polynomials $P_1,P_2$ belong to the same class if and only if $P_1-P_2$ is an MPI. These classes form a vector space $\Q_{D,m}$, the quotient space of $m$-degree homogeneous central polynomials by MPIs. Let $\{P_i(X)\}_i$ be a basis for $\Q_{D,m}$ and define the $m$-local operator $C\equiv\sum_{i}\ket{\varphi_i}\bra{P_i(X)}$, where $\{\ket{\varphi_i}\}_i$ is any orthonormal set of $m$-particle states. The effect of $C$ over any hMPS of bond dimension $D$ or smaller is to project the $m$ particles where it acts into the pure state $\ket{\varphi}\equiv\sum_i p_i(A)\ket{\varphi_i}$, while the remaining particles end up in the state (\ref{glued}). It hence `cuts' particles $s+1,...,s+m$ off the chain and `glues' the two ends back, see Figure \ref{cut_and_glue}.

This sort of operators will be called \emph{cut-and-glue operators}. Note that, after tracing out the particles cut, a cut-and-glue operator acting over an iMPS leaves the spin chain in the same quantum state (modulo normalization).

To understand the significance of cut-and-glue operators, consider this: given an injective iMPS $\ket{\psi}$, with matrices $A_{1},...,A_d$, one can always find homogeneous polynomials $\{P_i(X)\}_i$ of degree $m$ such that $P_i(A)\propto \id_D$ for all $i$ (this follows from the injectivity condition). Hence the operator $\sum_{i}\ket{i}\bra{P_i(X)}$ acting on $\ket{\psi}$ will produce a state of the form $(\sum_ip_i(A)\ket{i})\ket{\psi}$. However, that same operator applied over an arbitrary iMPS $\ket{\phi}$ with matrices $B_1,...,B_d$ will not produce a state proportional to $\ket{\phi}$, since $P_i(B)\not\propto \id_D$. Choosing $\{P_i(X)\}_i$ to be central polynomials, we make sure that $\sum_{i}\ket{i}\braket{P_i(X)}{\phi}=(\sum_ip_i(A)\ket{i})\ket{\phi}$ for \emph{all} iMPS $\ket{\phi}$ of bond dimension $D$.

Finally, note that cut-and-glue operators for dimension $D$ annihilate all non-injective iMPS of bond dimension $D$, as well as all iMPS with lower bond dimension. This follows from the fact that any central polynomial $P$ for dimension $D$ evaluated on matrices $A=(A_1,...,A_d)$ of size smaller than $D$ must equal zero. In effect, those can be embedded in $D\times D$ matrices as $\tilde{A}_i\equiv\left(\begin{array}{cc}A_i&0\\0&0\end{array}\right)$. Since by definition $P(\tilde{A})$ is proportional to the identity, the proportionality constant must be zero.

In Appendix \ref{CP_basis} we sketch efficient procedures to generate a basis for $\Q_{D,m}$. Table \ref{central} gives an idea of how the dimensionality of $\Q_{D,m}$ scales with the system size and the bond dimension in qubit ensembles. Surprisingly, it turns out that the dimensionality of $\Q_{2,m}$ in Table~\ref{central} follows the sequence of coefficients in the power series expansion of the Poincar\'e series $P(C_{2,2}; t)$, which is sequence A096338 in the \emph{On-Line Encyclopedia of Integer Sequences}~\cite{sloane}. We conjecture that the above observation holds true for entries beyond $m=15$ in Table~\ref{central} as well.

\begin{table}
\resizebox{\columnwidth}{!}{
\begin{tabular}{|c|c|c|c|c|c|c|c|c|c|c|c|c|c|}
\hline
$m$& 3&4&5&6&7&8&9&10&11&12&13&14&15\\
\hline
$D=2$& 0&1&2&6&10&20&30&50&70&105&140&196&252\\
$D=3$&0&0&0&0&0&0&4&16&50&129&274&542&$\times$\\
  \hline
\end{tabular}}
\caption{Dimensions of the $m$-qubit quotient spaces $\Q_{D,m}$ for different values of $m$ and bond dimensions $D=2,3$.}
\label{central}
\end{table}

Both annihilation and cut-and-glue operators are important structural features of hMPS. Unfortunately, even for $D=2$ their implementation in the lab would require the ability to switch on non-trivial four-interaction terms. This is experimentally challenging, given that in many experimental setups only 2-local operators are accessible. Fortunately, there is a cleverer way to exploit our findings. 

The notions of anihilation and cut-and-glue operators allow us to define a family of local operators $h$ whose average value is non-negative when computed with $n$-site MPS of bond dimension $D$ or smaller. Call $P$ the projector onto the space $\H^{MPS}_{D,n}$, and consider all $m$-local operators $h$ which satisfy:

\be
PhP=f+\sum_{j=1}^{n-1} PC_jg_jC_j^\dagger P,
\label{positive}
\ee

\noindent where $f\geq 0$, $C_j$ is a cut-and-glue operator acting non-trivially over particles $1,...,j$ and $g_j$ is an entanglement witness \cite{witness} with respect to the partition $1,...,j|j+1,...,n$ (namely, $\langle g_j\rangle\geq 0$ for all quantum states separable with respect to the said partition). Clearly, $\langle h\rangle_D\geq 0$. Note that, if we drop the requirement of locality, one can construct operators of the form (\ref{positive}) for arbitrarily high bond dimensions just by combining known families of entanglement witnesses, central polynomials and matrix polynomial identities.

Given an arbitrary ($k$-local) operator $H$, consider the problem of maximizing $\mu\in\R$ such that $H-\mu$ admits a decomposition of the form (\ref{positive}). Then, for any feasible $\mu$, $\langle H\rangle_D\geq \mu$. If the minimum eigenvalue of $H$ happens to be smaller than $\mu$, then $H-\mu$ can be regarded as a \emph{bond dimension witness}: an expectation value for $H$ below $\mu$ would prove that the underlying quantum state of the system does not admit a hMPS model of bond dimension $D$.

Regretfully, the optimization proposed above requires an implicit characterization of entanglement witnesses, a problem known to be NP-hard \cite{ent_complex}. A way out is to simply demand $g_j$ to belong to a class of entanglement witnesses which are easy to describe. An obvious choice is the set of all operators which are Positive (semidefinite) under Partial Transposition (PPT) \cite{peres}. With this restriction on $g_j$, the maximization of $\mu$ can be formulated as a semidefinite program (SDP), a class of convex optimization problems which can be solved efficiently \cite{sdp}. 

The dual of this program would be an optimization over all quantum states $\rho\in B(\H^{MPS}_{D,m})$ such that, for all $j$, $C_j\rho C_j^\dagger$ is PPT for the partition $1,...,j|j+1,...,n$. Since $\H^{MPS}_{D,m}$ grows polynomially with $m$, for small $D$ a normal computer can reach large values of $m$. Moreover, playing with the displacement operator, it is easy to derive a hierarchy of SDPs for the characterization of iMPS models. 

Rather than describing these tools in detail [the reader can find a full description of the SDP programs in Appendix \ref{SDP_rel}], we will illustrate how these methods work with a practical example. Consider an $N$-site spin 1/2 chain, and suppose that, via neutron interferometry, we estimate the expectation value of the XXX Heisenberg Hamiltonian 

\be
H_N=\sum_{i=1}^{N-1}\frac{1}{4}\vec{\sigma}_i\cdot\vec{\sigma}_{i+1}.
\ee

\noindent We wonder whether our experiment can be explained with a hMPS model of low bond dimension.  

Take the number of particles in the chain to be small, say $N=7$. For low $N$, the minimum eigenvalue of $H_N$ can be computed exactly, and so we find that the minimum average energy per interaction term is $E\equiv\min\frac{1}{6}\langle H_7\rangle\approx -0.4727$. On the other hand, an SDP optimization over 7-site normalized density matrices $\rho\in B(\H^{MPS}_{2,7})$ satisfying $C_j\rho C_j^\dagger$, PPT for $j=5,6$ returns the greater value $E_2\equiv\frac{1}{6}\langle H_7\rangle_2\geq -0.4065$. This optimization, and all subsequent ones, was carried out with the SDP solver MOSEK \cite{mosek}. The XXX Heisenberg Hamiltonian can thus be interpreted as a displaced bond dimension witness: an expectation value smaller than -0.4065, within reach given the lower value of $E$, would signify that the state of the spin chain cannot have a hMPS model of bond dimension $D=2$. 

Refuting $D=3$ hMPS models for $N=7$ is impossible with the tools developed so far, since $\H^{MPS}_{3,N}=(\C^2)^{\otimes N}$ and $\Q_{3,N}=\{0\}$, for $N=1,...,8$, see Tables \ref{MPS_dim} and \ref{central}. To make matters worse, SDP optimizations for $N\geq 9$ are too memory-demanding for a normal computer. However, if we drop the PPT condition, the resulting SDP can be seen equivalent to projecting $\frac{1}{N-1}H_N$ on the subspace $\H^{MPS}_{3,N}$ and finding the minimum eigenvalue of the resulting operator. This simplified method allows us to reach greater values of $N$, at the price of losing robustness in our bounds. With this trick, for $N=13$ we obtain a bound $E_3\equiv\frac{1}{12}\langle H_{13}\rangle_3\geq -0.44958$, slightly bigger than the minimum energy density $E=-0.46044$ achievable.

For $N\gg 1$, we can take the system to be approximately translational invariant, so this time we want to refute iMPS models of low bond dimension for our system. For $D=d=2$ iMPS models, it can be shown that spin chains are symmetric under parity, and hence satisfy non-trivial linear constraints. However, if our experimental setup does not allow us to estimate quantities of the sort $\sum_{i=1}^{N-1}\langle\sigma^s_{i}\sigma^t_{i+1}\rangle$, with $t\not=s$, we must again rely on inequalities, in which case the XXX Heisenberg Hamiltonian can also serve as a witness. It is a standard result (see, e.g., \cite{HeisXXX}) that the ground state energy density $E=\lim_{N\to\infty}\frac{1}{N-1}\langle H_N\rangle$ of the infinite XXX Heisenberg model is given by $E=1/4 - \ln(2)\approx -0.4431$. 

Now, how to derive bounds for iMPS models? A possibility would be to compute the minimum expectation value of $\frac{1}{N-1}\langle H_N\rangle$ for high $N$ via the hMPS SDP relaxation used above. Intuitively, increasing values of $N$ should give better and better approximations to the optimal iMPS value for the energy density. We chose, though, to use the slightly better approximation of optimizing the value of $\frac{1}{4}\vec{\sigma}_1\cdot\vec{\sigma}_{2}$ over reduced density matrices subject to the constraints above and the extra condition $\rho\in \S_{D,N}$, where $\S_{D,N}$ denotes the span of the $N$-site reduced density matrices of iMPS of bond dimension $D$. This space can be characterized using similar techniques as the ones we applied to compute $\H^{MPS}_{D,m}$. 

Optimizing over 8-site normalized reduced density matrices $\rho\in B(\H^{MPS}_{2,8})$ satisfying $C_j\rho C_j^\dagger$, PPT for $j=5,6,7$ and $\rho\in \S_{2,8}$ we find that $E_2\equiv\lim_{N\to\infty}\frac{1}{N-1}\langle H_N\rangle_2\geq -0.3378$. An average energy lower than the last value will hence refute all iMPS models of bond dimension $D=2$. 

This example is very illuminating in that it allows to appreciate the relevance of cut-and-glue operators for this class of optimizations. For, if we drop the PPT conditions above, the lower bound on $E_2$ output by the computer decreases to $-0.4246$. This is still bigger than $E$, and so it also defines a bond dimension witness. However, it is one order of magnitude less robust than the previous one, and so its violation is more challenging from an experimental point of view.

As a final example, we consider the Majumdar-Ghosh Hamiltonian \cite{MaGho}:

\be
H_{MG}= \sum_{i=1}^{N-2} \frac{1}{8}(2\vec{\sigma}_i\vec{\sigma}_{i+1}+\vec{\sigma}_i\vec{\sigma}_{i+2}).
\ee

\noindent The expectation value of this operator can also be estimated experimentally via neutron diffusion. In the thermodynamical limit $N\to\infty$, the minimum energy density of $H_{MG}$ is $E=-\frac{3}{8}=-0.375$, achievable with an iMPS of bond dimension $D=3$ \cite{mps}. In contrast, an $8^{th}$-order SDP relaxation over iMPS of bond dimension $D=2$ gives $E_2\geq-0.2593$. We have just derived a bond dimension witness with a large gap between iMPS models with $D=2$ and $D=3$. On the negative side, though, our lower bound for $E_2$ is significantly lower than the best upper bound $E_2\leq -0.125$ we found using the Amoeba variational method \cite{amoeba}.

Note that the former SDP methods can be easily turned into \emph{feasibility tests}. Indeed, determining the existence of a state with the properties above compatible with some partial information we may hold about the quantum state of the spin chain (such as, e.g., the average value of a number of $2$-local observables) can also be cast as an SDP. This procedure can help an experimentalist to refute the existence of a hMPS or iMPS model for the state he/she prepared in the lab, without the need of guessing the `right' bond dimension witness to do the job.

An immediate question is whether the SDP hierarchy of relaxations for iMPS models sketched above is complete, in the sense that it allows us to detect any state lacking an iMPS model by taking $N$ sufficiently large (the SDP relaxation for general hMPS is clearly \emph{not} complete). In this regard, notice that $\H^{MPS}_{1,N}$ corresponds to the symmetric space of $N$ particles, and $\Q_{1,N}=\H^{MPS}_{1,N}$. Hence for $D=1$ the hierarchy reduces to just imposing that the overall state is symmetric and PPT with respect to any bipartition. This is actually the Doherty-Parrilo-Spedalieri (DPS) method for entanglement detection \cite{DPS}, and convergence follows from the quantum de Finetti theorem \cite{finetti}. Similarly, $\H^{MPS}_{\infty,N}=\C^{d^N}$, $\Q_{\infty,N}=\{0\}$ and $\S_{\infty,N}$ is the span of all TI states. For $D=\infty$ the hierarchy is therefore computing the minimum expectation value of a Hamiltonian term over $N$-site states whose $N-1$-site reduced density matrices coincide whenever we remove the first or the last site. This is essentially a reformulation of Anderson's approximation \cite{anderson}, where convergence is also known to hold. One would be tempted to claim that our SDP hierarchy should converge as well for all intermediate values of $D$, but we leave this matter open.

\section{Applications for optimizations over finite dimensional quantum correlations}

In \cite{fin_dim,fin_dim_long}, a hierarchy of SDP relaxations is presented to characterize the statistics of finite-dimensional quantum systems. This hierarchy relies on the notion of \emph{moment matrices}. Given a quantum system in state $\sigma\in B(\C^D)$, with (self-adjoint) operators $X_1,...,X_{d-1}\subset B(\C^D)$, its $n^{th}$ order moment matrix $M$ is a matrix whose rows and columns are labeled by monomials $u$ of $X_1,...,X_{d-1}$ of degree smaller than or equal to $n$, with entries given by $M_{u,v}=\tr(u^\dagger(X) \sigma v(X))$. In \cite{fin_dim,fin_dim_long}, it is proposed to relax the requirement of $M$ admitting a quantum representation by demanding $M\geq 0$ and $M\in \M_D$, where $\M_D$ denotes the space spanned by moment matrices with quantum representations of dimension $D$.

A disadvantage of this method is that, for fixed $d,D$, the complexity of implementing the hierarchy increases exponentially with the index $n$ of the relaxation. In the following, we show that every $n^{th}$-order moment matrix with a quantum representation of dimension $D$ can be interpreted as a conic combination of $n$-site hMPS with bond dimension $D$. This will allow us to devise a hMPS-based algorithm that carries out an improved version of the $n^{th}$-order relaxation described in \cite{fin_dim,fin_dim_long} in time polynomial in $n$. 

Define $X_d\equiv\id_D$, and consider vectors of $d$ indices with values in $\{1,...,d\}$. Then, for any index vector $\vec{i}\in\{1,...,d\}^k$, we can associate the monomial $u(X)_{\vec{i}}\equiv X_{i_1}...X_{i_n}$. This procedure gives an over-representation of the set of monomials of degree smaller than or equal to $n$.

Now suppose that, by repeating rows and columns, we enlarge the $n^{th}$-order moment matrix $M$ of the system to an $d^k\times d^k$ matrix $\bar{M}$ such that $\bra{\vec{i}}\bar{M}\ket{\vec{j}}=M_{u_{\vec{i}},v_{\vec{j}}}$, with $\ket{\vec{i}}=\ket{i_1}...\ket{i_n}$ and similarly for $\ket{\vec{j}}$. The `enhanced' moment matrix $\bar{M}$ can then be written as

\begin{align}
\bar{M}_k=\sum_{\vec{i},\vec{j}}\tr(X_{i_n}^\dagger...X_{i_1}^\dagger\sigma X_{j_1}...X_{j_n})\ket{\vec{i}}\bra{\vec{j}}.
\label{connection_MPS}
\end{align}

\noindent This is just the transpose of a conic combination of hMPS of bond dimension $D$; much like eq. (\ref{TI_MPS}), but without the condition $\sum_i A^\dagger_i \sigma A_i=\sigma$. As such, its support is contained in $\H_{D,n}^{MPS}$; more precisely, in the analog set for hMPS with $X_d=\id$ plus any other extra restriction in the variables $X$. Hence $\bar{M}$ can be fully specified by a number of parameters polynomial in $n$.

Most interestingly, one can compute special cut-and-glue operators $C$ for this kind of hMPS. The convergence of the scheme can therefore be boosted by demanding extra positive semidefinite constraints such as $(C_j\bar{M} C_j^\dagger)^{T_j}\geq 0$. `Localizing matrices' of the form $M^q_{u,v}=\sum_{\vec{i},\vec{j}}\tr(X_{\vec{i}}^\dagger \sigma X_{\vec{j}} q(X))\ket{\vec{i}}\bra{\vec{j}}$, defined in \cite{fin_dim_long} to model semi-algebraic conditions of the sort $q(X)\geq 0$, can be treated in a similar way.

%This implies that $n^{th}$-order moment matrices live in a matrix space whose dimension scales just polynomially with the index $n$. Furthermore, the action of a cut-and-glue operator over a feasible moment matrix must return an operator which is positive under partial transposition. Exploiting these two observations, in Appendix \ref{complexity} we devise 

\section{Extension to Projected Entangled Pairs States}

\begin{figure}
  \centering
  \includegraphics[width=8 cm]{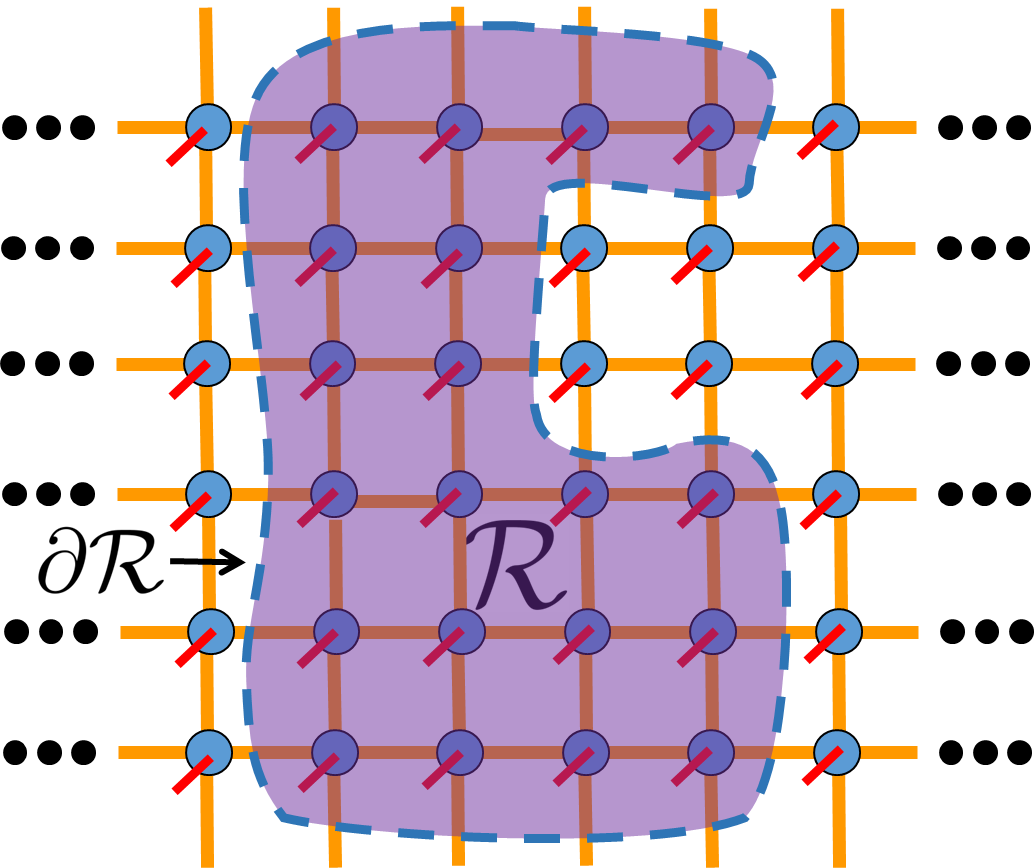}
  \caption{The region $\R$ in the two-dimensional lattice is marked in purple. Its boundary $\partial \R$ (dashed line) corresponds to all broken links of the repeated tensor.}
  \label{pepes}
\end{figure}

MPS can be understood as elements of a larger class of TNS called \emph{Projected Entangled Pairs States} (PEPS) \cite{peps}. Such states are used to approximate the low energy sector of local Hamiltonians describing particles sited in square lattices of arbitrary spatial dimensions. While MPS are defined via tensors with one physical index ($i=1,...,d$) and two bond indices (the column and row indices of the matrices $A_1,...,A_d$), $N$-dimensional PEPS are defined via contractions of tensors with 1 physical index and $2N$ bond indices. MPS can therefore be regarded as one-dimensional PEPS. For illustration, in Figure \ref{pepes} the tensors of a two-dimensional PEPS are represented by circles, while physical [bond] indices are denoted by red [orange] lines. If we assume that all such tensors are equal, we arrive at the notion of \emph{homogeneous PEPS} (hPEPS). It is natural to ask whether some of the structural features we derived for hMPS also extend to hPEPS of higher spatial dimension.

Consider the space spanned by hPEPS of bond dimension $D$ and physical dimension $d$ in a given region $\R$ of the lattice, with boundary $\partial \R$, see Figure \ref{pepes}. As with hMPS, we can express any hPEPS in $\R$ as in (\ref{mon_dec}), where each monomial $w$ has degree $1$ on the boundary condition and degree $|\R|$ on the tensor $A$ generating the hPEPS. The number of such monomials is $D^{|\partial\R|}\left(\begin{array}{c}|\R|+dD^2-1\\dD^2-1\end{array}\right)$, and hence, for fixed $d,D$, the local space spanned by hPEPS is upper bounded by $D^{|\partial\R|}\mbox{poly}(|\R|)$. This bound must be compared with the total dimensionality of the physical space in $\R$, namely, $d^{|\R|}$. Provided that $d^{|\R|}>\mbox{poly}(|\R|)D^{|\partial\R|}$, we will find non-trivial operators in $\R$ which will annihilate all hPEPS with bond dimension $D$ or smaller. If the spatial dimension of the lattice is $N$, taking $\R$ to be a hypercube of size $L$, with volume $|\R|=L^N$ and surface area $|\partial\R|=2NL^{N-1}$, this will happen for $L$ high enough.

We have just proven the existence of \emph{tensor polynomial identities}, i.e., local vectors $\ket{\phi}$ which annihilate all PEPS of bond dimension $D$. Given that generic hPEPS are the ground state of a unique parent Hamiltonian, it is easy to prove a weaker version of Proposition \ref{method_fails} for hPEPS of arbitrary spatial dimension. Namely, that the chain of identities will break at some point beyond $\langle H\rangle_D$ (not necessarily at $D+1$). Let us remark that, contrarily to hMPS variational algorithms, current tools for optimizations over hPEPS do not allow the user to reach high bond dimensions. Hence, even if the Hamiltonian is $k$-local, it may be that we can just compute the values $\{\langle H\rangle_D: D\ll k\}$. In such a predicament we may be eager to believe that the last estimation is a good approximation to the ground state energy, if the corresponding optimizations over lower bond dimensions returned similar results... and we could be wrong, as the arguments above show.

\section{Conclusion}

In this work, we have presented two highly non-trivial structural properties of hMPS, namely, the existence of annihilation and cut-and-glue operators. We used these notions to prove several results concerning the limitations of the hMPS approximation. Along the way, we raised a number of important open questions. 

First, it would be desirable to find closed formulas for the dimensionalities of $\H^{MPS}_{D,m}$ and $\Q_{D,m}$. Even though we have efficient methods to calculate these exactly, large values of $m$ require a considerable amount of computational time (hence the missing entries in Tables \ref{MPS_dim}, \ref{central}). Perhaps the connection with the Poincar\'e series $P(C_{2,2}; t)$ can be exploited in this regard.

Another interesting problem is whether our SDP hierarchy of relaxations to refute iMPS models is complete or can be further improved. 

Regarding completeness, for $D=1$ the proof of convergence follows from the quantum de Finetti theorem \cite{finetti}. For $D=\infty$, the hierarchy is just a reformulation of Anderson's approximation \cite{anderson}, whose convergence was established long ago. A convergence proof for all other values of $D$ would not only provide us with an alternative definition of iMPS, but most likely would involve an intermediate result of depth comparable to the quantum de Finetti theorem. 

As for improvement, a promising avenue to boost the speed of convergence of the hierarchy is to incorporate to our codes entropic constraints of the form $S(\rho_{1,...,k})\geq S(\rho_{1,...,k-1})$ for $2\leq k\leq n$, as in \cite{poulin}. These hold for any TI state; and, in particular, for iMPS. Although not reducible to SDPs, the corresponding problems can nonetheless be attacked with the tools of convex optimization theory. Considering the space spanned by \emph{several copies} of a hMPS should also help. 

Admittedly, it is difficult to believe that these ideas can ever lead to non-trivial restrictions for hMPS models with $D>4$. Devising new tools for the characterization of hMPS models of high bond dimension is hence an important matter.

Finally, it is intriguing whether the analogs of cut-and-glue operators for hMPS also exist for hPEPS of higher dimension. The action of such local operators over an arbitrary hPEPS would be to project the region where they act into a pure state and interconnect the links of the particles in the boundary. Appropriately tamed, such operators would allow transforming tensor network states of different type into each other by means of fixed (i.e., state-independent) local operations, very much like graph states transform into each other \cite{graph}. In this case, however, computational explorations face the exponential complexity of characterizing the space spanned by hPEPS of dimensions two and higher.

\vspace{10pt}
\noindent\emph{Acknowledgements}
We thank Andreas Winter for proving that the dimensionality of $\H^{MPS}_{D,m}$ grows polynomially with $m$. T.V. acknowledges the support of the National Research, Development and Innovation Office NKFIH (Grant Nos. K111734 and KH125096). This work was not funded by the European Research Council.

\bibliographystyle{plainnat}
\bibliography{bibQuantum}

\begin{thebibliography}{31}
\providecommand{\natexlab}[1]{#1}
\providecommand{\url}[1]{\texttt{#1}}
\expandafter\ifx\csname urlstyle\endcsname\relax
  \providecommand{\doi}[1]{doi: #1}\else
  \providecommand{\doi}{doi: \begingroup \urlstyle{rm}\Url}\fi

\bibitem[Aharonov et~al.(2009)Aharonov, Gottesman, Irani, and Kempe]{complex2}
Dorit Aharonov, Daniel Gottesman, Sandy Irani, and Julia Kempe.
\newblock The power of quantum systems on a line.
\newblock \emph{Communications in Mathematical Physics}, 287\penalty0
  (1):\penalty0 41--65, jan 2009.
\newblock \doi{10.1007/s00220-008-0710-3}.
\newblock URL \url{https://doi.org/10.1007%2Fs00220-008-0710-3}.

\bibitem[Anderson(1951)]{anderson}
P.~W. Anderson.
\newblock Limits on the energy of the antiferromagnetic ground state.
\newblock \emph{Physical Review}, 83\penalty0 (6):\penalty0 1260--1260, sep
  1951.
\newblock \doi{10.1103/physrev.83.1260}.
\newblock URL \url{https://doi.org/10.1103%2Fphysrev.83.1260}.

\bibitem[ApS(2015)]{mosek}
MOSEK ApS.
\newblock \emph{The MOSEK optimization toolbox for MATLAB manual. Version 7.1
  (Revision 28).}, 2015.
\newblock URL \url{http://docs.mosek.com/7.1/toolbox/index.html}.

\bibitem[Doherty et~al.(2002)Doherty, Parrilo, and Spedalieri]{DPS}
A.~C. Doherty, Pablo~A. Parrilo, and Federico~M. Spedalieri.
\newblock Distinguishing separable and entangled states.
\newblock \emph{Physical Review Letters}, 88\penalty0 (18), apr 2002.
\newblock \doi{10.1103/physrevlett.88.187904}.
\newblock URL \url{https://doi.org/10.1103%2Fphysrevlett.88.187904}.

\bibitem[Evenbly and Vidal(2013)]{MERAS}
Glen Evenbly and Guifre Vidal.
\newblock Quantum criticality with the multi-scale entanglement renormalization
  ansatz.
\newblock In \emph{Springer Series in Solid-State Sciences}, pages 99--130.
  Springer Berlin Heidelberg, 2013.
\newblock \doi{10.1007/978-3-642-35106-8_4}.
\newblock URL \url{https://doi.org/10.1007%2F978-3-642-35106-8_4}.

\bibitem[Fannes et~al.(1992)Fannes, Nachtergaele, and Werner]{iMPS2}
M.~Fannes, B.~Nachtergaele, and R.~F. Werner.
\newblock Finitely correlated states on quantum spin chains.
\newblock \emph{Communications in Mathematical Physics}, 144\penalty0
  (3):\penalty0 443--490, mar 1992.
\newblock \doi{10.1007/bf02099178}.
\newblock URL \url{https://doi.org/10.1007%2Fbf02099178}.

\bibitem[Formanek(1991)]{MPI}
Edward Formanek.
\newblock \emph{The Polynomial Identities and Variants of $n \times n$
  Matrices}.
\newblock American Mathematical Society, jan 1991.
\newblock \doi{10.1090/cbms/078}.
\newblock URL \url{https://doi.org/10.1090%2Fcbms%2F078}.

\bibitem[Gross et~al.(2007)Gross, Eisert, Schuch, and Perez-Garcia]{gross}
D.~Gross, J.~Eisert, N.~Schuch, and D.~Perez-Garcia.
\newblock Measurement-based quantum computation beyond the one-way model.
\newblock \emph{Physical Review A}, 76\penalty0 (5), nov 2007.
\newblock \doi{10.1103/physreva.76.052315}.
\newblock URL \url{https://doi.org/10.1103%2Fphysreva.76.052315}.

\bibitem[Gurvits(2004)]{ent_complex}
Leonid Gurvits.
\newblock Classical complexity and quantum entanglement.
\newblock \emph{Journal of Computer and System Sciences}, 69\penalty0
  (3):\penalty0 448--484, nov 2004.
\newblock \doi{10.1016/j.jcss.2004.06.003}.
\newblock URL \url{https://doi.org/10.1016%2Fj.jcss.2004.06.003}.

\bibitem[Hein et~al.(2004)Hein, Eisert, and Briegel]{graph}
M.~Hein, J.~Eisert, and H.~J. Briegel.
\newblock Multiparty entanglement in graph states.
\newblock \emph{Physical Review A}, 69\penalty0 (6), jun 2004.
\newblock \doi{10.1103/physreva.69.062311}.
\newblock URL \url{https://doi.org/10.1103%2Fphysreva.69.062311}.

\bibitem[Karbach et~al.(1998)Karbach, Hu, and Mu{\"u}ller]{HeisXXX}
Michael Karbach, Kun Hu, and Gerhard Mu{\"u}ller.
\newblock Introduction to the bethe ansatz {II}.
\newblock \emph{Computers in Physics}, 12\penalty0 (6):\penalty0 565, 1998.
\newblock \doi{10.1063/1.168740}.
\newblock URL \url{https://doi.org/10.1063%2F1.168740}.

\bibitem[König and Renner(2005)]{finetti}
Robert König and Renato Renner.
\newblock A de finetti representation for finite symmetric quantum states.
\newblock \emph{Journal of Mathematical Physics}, 46\penalty0 (12):\penalty0
  122108, dec 2005.
\newblock \doi{10.1063/1.2146188}.
\newblock URL \url{https://doi.org/10.1063%2F1.2146188}.

\bibitem[Levin and Nave(2007)]{iPEPS2}
Michael Levin and Cody~P. Nave.
\newblock Tensor renormalization group approach to two-dimensional classical
  lattice models.
\newblock \emph{Physical Review Letters}, 99\penalty0 (12), sep 2007.
\newblock \doi{10.1103/physrevlett.99.120601}.
\newblock URL \url{https://doi.org/10.1103%2Fphysrevlett.99.120601}.

\bibitem[Majumdar and Ghosh(1969)]{MaGho}
Chanchal~K. Majumdar and Dipan~K. Ghosh.
\newblock On next-nearest-neighbor interaction in linear chain. i.
\newblock \emph{Journal of Mathematical Physics}, 10\penalty0 (8):\penalty0
  1388--1398, aug 1969.
\newblock \doi{10.1063/1.1664978}.
\newblock URL \url{https://doi.org/10.1063%2F1.1664978}.

\bibitem[Navascu{\'{e}}s and V{\'{e}}rtesi(2015)]{fin_dim}
Miguel Navascu{\'{e}}s and Tam{\'{a}}s V{\'{e}}rtesi.
\newblock Bounding the set of finite dimensional quantum correlations.
\newblock \emph{Physical Review Letters}, 115\penalty0 (2), jul 2015.
\newblock \doi{10.1103/physrevlett.115.020501}.
\newblock URL \url{https://doi.org/10.1103%2Fphysrevlett.115.020501}.

\bibitem[Navascu{\'{e}}s et~al.(2015)Navascu{\'{e}}s, Feix, Ara{\'{u}}jo, and
  V{\'{e}}rtesi]{fin_dim_long}
Miguel Navascu{\'{e}}s, Adrien Feix, Mateus Ara{\'{u}}jo, and Tam{\'{a}}s
  V{\'{e}}rtesi.
\newblock Characterizing finite-dimensional quantum behavior.
\newblock \emph{Physical Review A}, 92\penalty0 (4), oct 2015.
\newblock \doi{10.1103/physreva.92.042117}.
\newblock URL \url{https://doi.org/10.1103%2Fphysreva.92.042117}.

\bibitem[Oliveira and Terhal(2008)]{complex1}
Roberto Oliveira and Barbara~M. Terhal.
\newblock The complexity of quantum spin systems on a two-dimensional square
  lattice.
\newblock \emph{Quant. Inf, Comp.}, 8, 2008.

\bibitem[Or{\'{u}}s(2014)]{tensor_net}
Rom{\'{a}}n Or{\'{u}}s.
\newblock A practical introduction to tensor networks: Matrix product states
  and projected entangled pair states.
\newblock \emph{Annals of Physics}, 349:\penalty0 117--158, oct 2014.
\newblock \doi{10.1016/j.aop.2014.06.013}.
\newblock URL \url{https://doi.org/10.1016%2Fj.aop.2014.06.013}.

\bibitem[Peres(1996)]{peres}
Asher Peres.
\newblock Separability criterion for density matrices.
\newblock \emph{Physical Review Letters}, 77\penalty0 (8):\penalty0 1413--1415,
  aug 1996.
\newblock \doi{10.1103/physrevlett.77.1413}.
\newblock URL \url{https://doi.org/10.1103%2Fphysrevlett.77.1413}.

\bibitem[Perez-Garc\'ia et~al.(2007)Perez-Garc\'ia, Verstraete, Wolf, and
  Cirac]{mps}
D.~Perez-Garc\'ia, F.~Verstraete, M.~M. Wolf, and J.I. Cirac.
\newblock Matrix product state representations.
\newblock \emph{Quantum Inf. Comput.}, 7:\penalty0 401, sep 2007.

\bibitem[Phien et~al.(2015)Phien, Bengua, Tuan, Corboz, and Or{\'{u}}s]{iPEPS}
Ho~N. Phien, Johann~A. Bengua, Hoang~D. Tuan, Philippe Corboz, and Rom{\'{a}}n
  Or{\'{u}}s.
\newblock Infinite projected entangled pair states algorithm improved: Fast
  full update and gauge fixing.
\newblock \emph{Physical Review B}, 92\penalty0 (3), jul 2015.
\newblock \doi{10.1103/physrevb.92.035142}.
\newblock URL \url{https://doi.org/10.1103%2Fphysrevb.92.035142}.

\bibitem[Poulin and Hastings(2011)]{poulin}
David Poulin and Matthew~B. Hastings.
\newblock Markov entropy decomposition: A variational dual for quantum belief
  propagation.
\newblock \emph{Physical Review Letters}, 106\penalty0 (8), feb 2011.
\newblock \doi{10.1103/physrevlett.106.080403}.
\newblock URL \url{https://doi.org/10.1103%2Fphysrevlett.106.080403}.

\bibitem[Schuch et~al.(2010)Schuch, Cirac, and P{\'{e}}rez-Garc{\'{\i}}a]{peps}
Norbert Schuch, Ignacio Cirac, and David P{\'{e}}rez-Garc{\'{\i}}a.
\newblock {PEPS} as ground states: Degeneracy and topology.
\newblock \emph{Annals of Physics}, 325\penalty0 (10):\penalty0 2153--2192, oct
  2010.
\newblock \doi{10.1016/j.aop.2010.05.008}.
\newblock URL \url{https://doi.org/10.1016%2Fj.aop.2010.05.008}.

\bibitem[Sloane()]{sloane}
Neil J.~A. Sloane.
\newblock The on-line encyclopedia of integer sequences.
\newblock In \emph{Towards Mechanized Mathematical Assistants}, pages 130--130.
  Springer Berlin Heidelberg.
\newblock \doi{10.1007/978-3-540-73086-6_12}.
\newblock URL \url{https://doi.org/10.1007%2F978-3-540-73086-6_12}.

\bibitem[Östlund and Rommer(1995)]{iMPS1}
Stellan Östlund and Stefan Rommer.
\newblock Thermodynamic limit of density matrix renormalization.
\newblock \emph{Physical Review Letters}, 75\penalty0 (19):\penalty0
  3537--3540, nov 1995.
\newblock \doi{10.1103/physrevlett.75.3537}.
\newblock URL \url{https://doi.org/10.1103%2Fphysrevlett.75.3537}.

\bibitem[Terhal(2000)]{witness}
Barbara~M. Terhal.
\newblock Bell inequalities and the separability criterion.
\newblock \emph{Physics Letters A}, 271\penalty0 (5-6):\penalty0 319--326, jul
  2000.
\newblock \doi{10.1016/s0375-9601(00)00401-1}.
\newblock URL \url{https://doi.org/10.1016%2Fs0375-9601%2800%2900401-1}.

\bibitem[Vandenberghe and Boyd(1996)]{sdp}
Lieven Vandenberghe and Stephen Boyd.
\newblock Semidefinite programming.
\newblock \emph{{SIAM} Review}, 38\penalty0 (1):\penalty0 49--95, mar 1996.
\newblock \doi{10.1137/1038003}.
\newblock URL \url{https://doi.org/10.1137%2F1038003}.

\bibitem[Verstraete et~al.(2004)Verstraete, Garc{\'{\i}}a-Ripoll, and
  Cirac]{MPO}
F.~Verstraete, J.~J. Garc{\'{\i}}a-Ripoll, and J.~I. Cirac.
\newblock Matrix product density operators: Simulation of finite-temperature
  and dissipative systems.
\newblock \emph{Physical Review Letters}, 93\penalty0 (20), nov 2004.
\newblock \doi{10.1103/physrevlett.93.207204}.
\newblock URL \url{https://doi.org/10.1103%2Fphysrevlett.93.207204}.

\bibitem[Verstraete et~al.(2008)Verstraete, Murg, and Cirac]{app_MPS}
F.~Verstraete, V.~Murg, and J.I. Cirac.
\newblock Matrix product states, projected entangled pair states, and
  variational renormalization group methods for quantum spin systems.
\newblock \emph{Advances in Physics}, 57\penalty0 (2):\penalty0 143--224, mar
  2008.
\newblock \doi{10.1080/14789940801912366}.
\newblock URL \url{https://doi.org/10.1080%2F14789940801912366}.

\bibitem[Werner(2006)]{Werner_en}
R.F. Werner.
\newblock Finitely correlated states.
\newblock In \emph{Encyclopedia of Mathematical Physics}, pages 334--340.
  Elsevier, 2006.
\newblock \doi{10.1016/b0-12-512666-2/00379-5}.
\newblock URL \url{https://doi.org/10.1016%2Fb0-12-512666-2%2F00379-5}.

\bibitem[Ziegel et~al.(1987)Ziegel, Press, Flannery, Teukolsky, and
  Vetterling]{amoeba}
Eric Ziegel, William Press, Brian Flannery, Saul Teukolsky, and William
  Vetterling.
\newblock Numerical recipes: The art of scientific computing.
\newblock \emph{Technometrics}, 29\penalty0 (4):\penalty0 501, nov 1987.
\newblock \doi{10.2307/1269484}.
\newblock URL \url{https://doi.org/10.2307%2F1269484}.

\end{thebibliography}

\begin{appendix}
\section{Exploring $\H_{D,m}^{MPS}$}
\label{MPS_basis}

Except for $D=1$, the upper bound defined by eq. (\ref{UB}) in the main text is not tight, the reason being that the vectors $\{\ket{\phi_w}\}_w$ are linearly dependent. This is (partly) due to the fact that different values of $A,\omega$ may correspond to the same hMPS. Indeed, take $D=d=2$, and note that, for any matrix $S\in B(\C^D)$, the transformation $\omega\to S^{-1}\omega S$, $A_i\to S^{-1}A_i S$ leaves the state in eq. (\ref{MPS}) of the main text invariant. In particular, we can choose $S$ to diagonalize $A_1$. Applying furthermore a diagonal transformation $T=\mbox{diag}(a,b)$ we can enforce that the off-diagonal elements of $A_2$ are equal. With this parametrization, $A$ is specified by just $5$ parameters, and so the dimensionality of $\H^{MPS}_{2,m}$ is expected to grow as $O(m^4)$, rather than $O(m^7)$, as the formula (\ref{UB}) in the text suggests.

Determining the exact dimensionality of $\H^{MPS}_{D,m}$ and deriving an orthonormal basis for this subspace can be done via two different procedures. First, given an arbitrary weight $w(\omega,A)>0$, we will call \emph{polynomial MPS} any $m$-site states of the form

\be
\ket{f^{MPS}}\equiv\int d\omega dA w(\omega,A) f(\omega,A)^*\ket{\psi(\omega,A,m)},
\ee

\noindent where $f(\omega,A)$ is a homogeneous polynomial of the components of $\omega$ (with degree 1) and $A$ (with degree $m$). A comfortable possibility is to take $\omega,A$ real and the trivial weight $w(\omega,A)=1$. Defining $t\equiv \dim(\H^{MPS}_{D,m})$, our task is to find a set of polynomials $\{f_i\}_{i=1}^t$ such that $\braket{f^{MPS}_i}{f^{MPS}_j}=\delta_{ij}$ and $\sp\{\ket{f_i^{MPS}}\}=\H^{MPS}_{D,m}$. This can be seen equivalent to diagonalizing the kernel

\begin{align}
&K(A,\omega,A',\omega')\equiv \nonumber\\
&\braket{\psi(\omega',A')}{\psi(\omega,A)}w(\omega,A)w(\omega',A')=\nonumber\\
&=\tr\left\{\left(\sum_{i=1}^d \bar{A}'_i\otimes A_i\right)^N\left(\bar{\sigma'}\otimes\sigma\right)\right\}w(\omega,A)w(\omega',A'),
\end{align}

\noindent on a basis of homogeneous polynomials of degree $1$ in $\omega$ and $m$ in the entries of $A$ and taking the eigenvectors (polynomials) with non-zero eigenvalue.

Alternatively, we can simply sequentially generate real random $D\times D$ matrices $\omega^j, A^j_1,...,A^j_d$ and use them to define the sequence of random hMPS $(\ket{\psi(\omega^j,A^j,m)})_j$. Exploiting the fact that the overlap between two hMPS can be computed efficiently, one can apply a Gram-Schmidt process to the previous sequence of hMPS, thus obtaining an orthonormal basis for $\H^{MPS}_{D,m}$, whose elements are finite linear combinations of hMPS.

\section{Proof of Proposition 1}
\label{proof_prop1}
The goal of this section is to prove Proposition 1 in the main text, which reads:

\begin{prop}
Let $D>D'>1$ be natural numbers. Then, for any $N$, there exists an $O(D^2)$-local TI $n$-qubit Hamiltonian $H$, with $n>N$, satisfying

\begin{align}
&\langle H\rangle_{D'-1}>\langle H\rangle_{D'}=\langle H\rangle_{D'+1}=...\nonumber\\
&...=\langle H\rangle_{D-1}=\langle H\rangle_{D}> \langle H\rangle_{D+1}.
\end{align}

\end{prop}

For the proof we need two intermediate results, namely:

\begin{enumerate}
\item There exist TI $O(D)$-local Hamiltonians of arbitrarily many particles whose unique ground state is a hMPS of bond dimension $D$. This is proven in Section \ref{parents}.
\item For any $D>1$, there exists a bivariate noncommutative homogeneous polynomial $F(X_1,X_2)$ of degree $O(D^2)$ that is a MPI for matrices of size $D-1\times D-1$, but not for matrices of size $D\times D$. This will be proven in Section \ref{sep_MPIs}.
\end{enumerate}

These two results will be combined to demonstrate the Proposition in Section \ref{proof_prop}.

\subsection{Parent Hamiltonians and MPIs}
\label{parents}

A set of matrices $A_1,...,A_d\in B(\C^D)$ satisfies the \emph{injectivity condition} if there exists $k$ such that the products $\{A_{i_1}...A_{i_k}\}$ span all of $B(\C^D)$. From \cite{mps}, we know that any $n$-site uniform TI MPS $\ket{\psi}$ whose matrices satisfy the injectivity condition for some order $k$ can be seen as the unique ground state of a (TI) $2k$-local Hamiltonian (provided that $n\geq 2k$). The latter is called the \emph{parent Hamiltonian} of $\ket{\psi}$.

We will next prove that, for any $D$, there exist matrices $B_1,B_2\in B(\C^D)$ which satisfy the injectivity condition for $k=O(D)$. With the above, this will imply that, for any bond and physical dimensions $D,d$ and any system size $n$, there exists an $n$-site TI MPS which arises as the unique ground state of a $O(D)$-local TI Hamiltonian.

Let $d=2$, and consider the matrices

\be
B_1=\sum_{j=1}^{D}j\proj{j}, B_2=\frac{1}{D}\sum_{i,j=1}^{D}\ket{i}\bra{j}
\label{injective}
\ee

\noindent Note that we can express the projectors $\{\proj{i}\}_{i=1}^D$ as linear combinations of $\{B_1^p:p=1,...,D\}$. Since $B_2^s=B_2$ for any $s\geq 1$, this implies that linear combinations of the (degree $2D+1$) products $B_1^{p} B_2^{2D-p-q+1} B_1^q$ can generate the matrices $\proj{i}B_2\proj{j}=\frac{1}{D}\ket{i}\bra{j}$, which span $B(\C^D)$.

\subsection{MPIs for dimension $D-1$ which cease to be identities in dimension $D$}
\label{sep_MPIs}

In this section we will prove that, for any $D$, there exists a polynomial $F(X_1,X_2)$ of degree $O(D^2)$ which is a MPI for dimension $D-1$, but not for dimension $D$.

Choose $B_1,B_2\in B(\C^{D})$ as in (\ref{injective}). As proven in Appendix \ref{parents}, there exist homogeneous polynomials $\{P_j(X)\}_j$ of degree $O(D)$ such that $P_1(B)=\proj{1},P_2(B)=\ket{1}\bra{2},...,P_{2D-3}(B)=\proj{D-1},P_{2D-2}(B)=\ket{D-1}\bra{D}$. Note that the matrices $\{P_j(B)\}_j$ have the peculiarity that the only permutation of them which does not vanish is $P_1(B)P_2(B)...P_{2(D-1)}(B)$. Hence, by construction, the standard polynomial [eq. \ref{funda} in the main text] with $N=2(D-1)$, applied to the tuple $Y=(P_1(B),P_2(B),...)$, results in a non-zero value. That is, the $O(D^2)$-degree homogeneous polynomial $P(X_1,X_2)\equiv F_{2(D-1)}(P_1(X),...,P_{2(D-1)}(X))$, while being an MPI for $D-1\times D-1$ matrices, is \emph{not} an MPI for dimension $D$.

\subsection{Putting all together}
\label{proof_prop}
Now we are ready to prove Proposition 1.

\begin{proof}

Take a TI $n$-qubit MPS $\ket{\psi}$ with bond dimension $D'$, and build its TI parent Hamiltonian $H$ \cite{mps}. Such is a $k$-local operator with the properties $H\geq 0$, $H\ket{\psi}=0$ and $\ket{\psi}$ being the only ground state of $H$. From Section \ref{parents} we know that $\ket{\psi}$ can be chosen injective and such that $H$ has interaction strength $k=O(D')$. Since $\ket{\psi}$ is injective, it cannot be expressed as a hMPS of bond dimension $D'-1$ (because, e.g., $\ket{\psi}$ has Schmidt rank greater than $D'-1$). It follows that $\langle H\rangle_{D'-1}>\langle H\rangle_{D'}=0$.

Now, for $m$ high enough, choose $h\in B(\H^{MPI}_{D,m})$, $h>0$, with $h\not\in B(\H^{MPI}_{D+1,m})$. From Section \ref{sep_MPIs} we know that, no matter the value of the physical dimension, there exists such an operator with $m=O(D^2)$. Given $h$, define the family of $O(D^2)$-local TI Hamiltonians $H_\lambda=H-\lambda \sum_{i=1}^n\tau^i(h)$, where $\tau$ is the translation operator. By construction, $H_\lambda\ket{\psi}= H\ket{\psi}$ for all hMPS of bond dimension smaller than or equal to $D$, and so the equalities in eq. (\ref{curioso}) are satisfied. On the other hand, for $\lambda$ high enough, $\langle H_\lambda\rangle_{D+1}<0$.

\end{proof}

Note that, if we are entitled to play with the physical dimension of the system, we do not need to invoke Section \ref{sep_MPIs} at all. Indeed, it suffices to set $d=2D$ and take $h$ to be the standard identity for dimension $D$ (which, having degree $2D$, cannot be an MPI for dimension $D+1$). The corresponding family of Hamiltonians $H_\lambda$ would then be $O(D)$-local, rather than $O(D^2)$-local.

\section{Characterizing $\Q_{D,m}$}
\label{CP_basis}

Viewed as a subspace of $(\C^d)^{\otimes m}$, the space of central polynomials $\H^{CP}_{D,m}$ corresponds to the orthogonal complement of

\begin{align}
&\H^{MPS,[,]}_{D,m}=\sp\{\sum_{i_1,...,i_m}\tr(\omega [B,A_{i_1},...,A_{i_m}])\ket{i_1,...,i_m}\}=\nonumber\\
&=\sp\{\sum_{i_1,...,i_n}\tr([\omega,B] A_{i_1},...,A_{i_m})\ket{i_1,...,i_m}\}=\nonumber\\
&=\sp\{\ket{\psi(\sigma,A,m)}:\tr(\sigma)=0\},
\end{align}

\noindent where the last equality follows from the equivalence between traceless matrices and commutators. Now, the quotient space $\Q_{D,m}$ corresponds to $\H^{CP}_{D,m}\cap (\H^{MPI}_{D,m})^{\perp}=\H^{CP}_{D,m}\cap \H^{MPS}_{D,m}$. A basis for $\Q_{D,m}$, orthonormal as a subspace of $(\C^d)^{\otimes m}$ can thus be obtained via the following procedure: first, we sequentially generate real random matrices $\sigma^j,A_1^j,...,A_d^j$, with $\tr(\sigma^j)=0$, which we use to construct a random basis $\{\ket{\phi^{[,]}_i}\}_i$ for $\H^{MPS,[,]}_{D,m}$ (just as we built a basis for $\H^{MPS}_{D,m}$ in Appendix \ref{MPS_basis}). Next, we find its orthogonal complement with respect to $\H^{MPS}_{D,m}$, the space of hMPS. This can be done, e.g., by determining the kernel of the matrix $A_{ij}\equiv\braket{\phi^{[,]}_i}{\phi_j}$, where $\{\ket{\phi_j}\}_j$ is a basis for $\H^{MPS}_{D,m}$.

As before, instead of using a randomized algorithm, we can find an orthonormal basis for $\H^{MPS,[,]}_{D,m}$ via polynomial hMPS $\{\ket{f_i}\}_i$ and use it to construct the matrix $\tilde{A}_{ij}=\braket{f_i}{g_j}$, where $\{\ket{g_j}\}_j$ denotes an orthonormal basis of polynomials for $\H^{MPS}_{D,m}$. The kernel of $\tilde{A}$ will give us a polynomial basis for $\Q_{D,m}$.

In either case, the dimensionality of all these spaces grows polynomially with the system size $m$, so for small $D$ we can find an orthonormal basis for $\H^{CP}_{D,m}$ for very high values of $m$.

\section{An SDP relaxation for linear optimizations over MPS}
\label{SDP_rel}

Given an $m$-site Hamiltonian $H$, consider the problem

\begin{align}
&\min \tr (H\rho),\nonumber\\
\mbox{s.t. } & \rho\in B(\H^{MPS}_{D,m}),\rho\geq 0,\tr(\rho)=1,\nonumber\\
&(C_j\rho C_j^\dagger)^{T_j}\geq 0,j=1,...,m-1
\label{sdp}
\end{align}

\noindent where $C_j$ is a cut-and-glue operator acting non-trivially over particles $1,...,j$ and $B^{T_{j}}$ denotes the partial transposition of $B$ with respect to the same particle set \cite{peres}. Note that we have relaxed the condition of $C_j\rho C_j^\dagger$ being separable to the simpler constraint of being positive under partial transposition. Clearly, the solution of the SDP (\ref{sdp}) will provide a \emph{lower bound} for $\langle H\rangle_D$. Let us remark that an implementation of (\ref{sdp}) requires explicit bases for $\H^{MPS}_{D,m}$ and $\{\Q_{D,j}:j=1,...,m-1\}$. These can be obtained efficiently with the non-deterministic algorithms described in Sections \ref{MPS_basis} and \ref{CP_basis}.

Suppose now that we are interested in optimizing over TI hMPS given by eq. (3) in the main text. In that case, one can define the following \emph{hierarchy} of SDPs:

\begin{align}
h^n\equiv &\min \frac{1}{n-m}\tr \{\sum_{i=1}^{n-m}\tau_i(H)\rho\},\nonumber\\
\mbox{s.t. } & \rho\in B(\H^{MPS}_{D,n}),\rho\geq 0,\tr(\rho)=1,\nonumber\\
&(C_j\rho C_j^\dagger)^{T_j}\geq 0,j=1,...,n-1.
\label{SDP}
\end{align}

\noindent Here $\tau$ denotes the translation operator, and is used to enforce TI on the relaxation. Another way to enforce TI is to determine (via., e.g., randomization) the span $\S_{D,n}$ of the $n$-site density matrices of all uniform TI MPS and then demand $\rho\in\S_{D,n}$. In either case, $h^m\leq h^{m+1}\leq...\leq h^\star$, where $h^\star$ is the minimum average value of $H$ over extendible MPS with bond dimension $D$.

For $D=1$, the space $\H^{MPS}_{D,n}$ reduces to the symmetric space of $n$ $d$-dimensional particles, and $C_j=\id$ (since all polynomials are central in $D=1$). The method hence reduces to the Doherty-Parrilo-Spedalieri (DPS) method for entanglement detection \cite{DPS}. The DPS method approximates the set of states of the form $\int d\phi p(\phi) \proj{\phi}^{\otimes m}$, with $p(\phi) \geq 0$ by partial traces of the set of $n$-symmetric states positive under partial transposition. It can be shown to converge by virtue of the quantum de-Finetti theorem \cite{finetti}. For $D=\infty$, program (\ref{SDP}) is equivalent to computing the $(n-m)^{th}$ Anderson bound for the Hamiltonian $H$ \cite{anderson}, and convergence can be proven easily. It is an open question whether the hierarchy (\ref{SDP}) converges for other values of $D$.

\end{appendix}

\end{document}